\documentclass[aps,pre,twocolumn,groupedaddress,showpacs]{revtex4}
\newif\ifpdf            
\ifx\pdfoutput\undefined \pdffalse \else \pdftrue \fi
\ifpdf 
\usepackage[pdftex]{graphicx}
\pdfcompresslevel9 \DeclareGraphicsExtensions{.jpg,.pdf,.png,.mps}
\else 
\usepackage[dvips]{graphicx}
\DeclareGraphicsExtensions{.eps,.ps}
\fi
\usepackage{bm}

\begin{document}

\title{Stereo-selective swelling of imprinted cholesteric networks}

\author{S.~Courty}
\author{A.R.~Tajbakhsh}
\author{E.M.~Terentjev}

\affiliation{Cavendish Laboratory, University of Cambridge,
Madingley Road, Cambridge, CB3 0HE, U.K. }

\date{\today}

\begin{abstract}
Molecular chirality, and the chiral symmetry breaking of resulting
macroscopic phases, can be topologically imprinted and manipulated
by crosslinking and swelling of polymer networks. We present a new
experimental approach to stereo-specific separation of chiral
isomers by using a cholesteric elastomer in which a helical
director distribution has been topological imprinted by
crosslinking. This makes the material unusual in that is has a
strong phase chirality, but no molecular chirality at all; we
study the nature and parameters controlling the twist-untwist
transition. Adding a racemic mixture to the imprinted network
results in selective swelling by only the component of ``correct''
handedness. We investigate the capacity of demixing in a racemic
environment, which depends on network parameters and the
underlying nematic order.
\end{abstract}


\maketitle

Due to the lack of inversion symmetry, chiral molecular objects
are non-superimposable with their mirror image. A pair of
opposite, left- and right-handed, isomers differs in shape only in
a very subtle way, which results in small corrections to
higher-order molecular polarizability. Although chirality was
known as a phenomenon since the middle of the 19-century, the
quantitative characterization of such a delicate physical property
has been only achieved in the last decade
\cite{Osipov95,Harris99}. It is now understood that handedness is
not an absolute concept but depends on the property being
observed; different scalar parameters describing chirality can
have differing signs and can vanish at different points as an
object is continuously deformed into its mirror image.

Chirality is a ubiquitous feature of the biological world. Many,
if not most functions of fundamental living components rely upon
their chirality. Thus the synthesis of a single enantiomer of a
compound, or its separation from a 50/50 racemic mixture, is
crucial for many practical applications -- from pharmaceutical, to
food and cosmetics. This stereo-selection is always a difficult
task, because the molecular interactions that are sensitive to its
handedness are always very small. One of the main techniques in
this field is column chromatography, in which a racemic mixture
diffuses through a silica gel coated with a molecular layer of
specific chirality: the two enantiomers of the mixture diffuse at
slightly different rates due to the weak (higher order) van der
Waals attraction to the gel coating.

Microscopic chiral constituents have a profound effect on the
macroscopic structures they form, striking examples of which are
common in liquid crystals \cite{deGennes93}. The simplest liquid
crystalline phase is the nematic, characterized by a long-range
uniaxial orientational order of rod-like anisotropic molecules
along a common director $\bm{n}$. Adding chiral molecules results
in a cholesteric phase, with the director twisting in space in a
periodic helical fashion, with the pitch inversely proportional to
the concentration of chiral dopant. Analogous states with
macroscopic ``phase chirality'' are found in blue phases and a
variety of smectic liquid crystals. The coherent arrangement of
molecular units on a macroscopic scale, such as the cholesteric
helix, greatly enhances physical manifestations of chirality. For
instance, the optical rotation (Faraday effect) of a solution of
chiral molecules can be enhanced by up to $10^4$ in a cholesteric
phase generated by the same amount of chiral doping. It is
attractive to utilize this macroscopic enhancement of normally
weak chiral interactions.

In this Letter, we demonstrate the effect of stereo-selection
between left- and right-handed molecules of a racemic mixture,
achieved in a polymer gel network in which cholesteric order has
been topologically imprinted by crosslinking.  Cholesteric
polymers and their networks have been known for a long time
\cite{Finkelmann:81,Freidzon:86,Zentel:87,Meier:90}, however, the
topological imprinting of phase chirality is a completely
different concept. The network crosslinked in the cholesteric
phase induced by a chiral dopant would retain a memory of
chirality even when the dopant is completely removed, leaving an
internally stored helical twist in a material with no chemical
(microscopic) chirality.  de Gennes was the first to suggest that
chiral order can be induced by crosslinking a conventional polymer
in a chiral liquid crystal phase \cite{deGennes:69a}.
Experimentally, elements of chiral imprinting have been
demonstrated in different polymer systems
\cite{Tsutsui:80,Hasson:98}. A continuum theory for chiral
imprinting in nematic elastomers formed in a chiral environment
has been developed by Mao and Warner (MW) \cite{Mao:00}, followed
by their theoretical suggestion about stereo-selective swelling
and chiral separation \cite{Mao:01b}. Qualitatively, the concept
is simple: chiral dopant molecules leave a specific ``trace'',
frozen-in by chemical crosslinking of polymer chains; on bringing
in a racemic solvent, the component of ``correct'' handedness
recognizes the trace and thus has a preference in swelling the
network. However, the real problem is highly non-trivial and
influenced by several competing factors. Our task here is to test
the basic theoretical model and explore stereo-selective
interactions according to the handedness of the imprinted network.

\begin{figure}
\resizebox{0.34\textwidth}{!}{\includegraphics{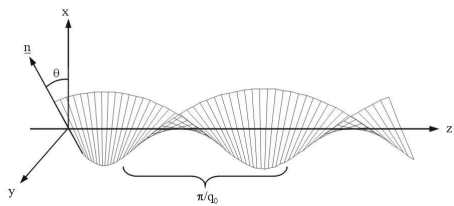}}
\caption{Spatial distribution of the director $\bm{n}$ in an ideal
cholesteric helix along the macroscopic optical axis $z$.}
\label{helix}
\end{figure}
Locally cholesterics are an amorphous uniaxial medium, described
by the local nematic order parameter $Q_{ij}= Q(T,\phi)
(n_{i}n_{j}-\frac{1}{3}\delta_{ij})$, with the director $\bm{n}$ a
periodic modulated function of coordinates, in the ideal state
rotating along a single axis $z$: $n_{x}=\cos\theta,
n_{y}=\sin\theta, n_{z}=0$. In the ideal cholesteric the azimuthal
angle is $\theta=q_{0}z$, with the corresponding helical pitch
$p_{0}= \pi/q_{0}$, Fig.~\ref{helix}. This spontaneously twisted
director distribution can be due to the presence of chiral
molecules in the nematic polymer network during its crosslinking.
If, after crosslinking, the chiral dopant is removed from this
network (or replaced by an achiral solvent), two competing
processes occur: The Frank energy penalty for the director twist,
$\frac{1}{2}K_{2} (\bm{n}\cdot curl \bm{n})^{2}$, demands the
cholesteric helix to unwind. The local anchoring of the director
to the rubbery network, measured by the relative-rotation coupling
constant $D_{1}$ \cite{deGennes80}, resists any director
rotations, thus acting to preserve the imprinted helix. $D_1$ is
proportional to the rubber modulus of the network and, through it,
to the crosslinking density \cite{OUP}. The free density energy is
then:
\begin{equation}
F = \int {\textstyle{\frac{1}{2}}} \left[ K_{2}
({\textstyle{\frac{d}{dz}}}\theta - q)^{2} +D_{1}\sin^{2} (\theta-
q_{0}z) \right] dz  \label{free energy}
\end{equation}
with $q$ the helical wave number that the current concentration of
chiral solvent $\phi$ would induce, $q=4\pi\beta\phi$, where
$\beta$ is the microscopic twisting power of the solute
\cite{deGennes93}. With its complete removal, $\phi=0$ and $q=0$,
while the concentration at crosslinking is taken as $\phi_0$, with
$q=q_0$. MW have quantified the balance between these two physical
trends by introducing a chiral order parameter $\alpha$:
\begin{equation}
\alpha= \xi [q_{0}-q(\phi)] \, , \ \ \ \ {\rm with} \ \ \xi
=\sqrt{K_{2}/D_{1}}  \label{alpha}
\end{equation}
the nematic rubber penetration length. Note that both $K_2$ and
$D_1$ are proportional to the square of local nematic parameter
$Q$, and so the length $\xi\approx \, $const. The resulting
classical problem of elliptical functions predicts that the helix
coarsens and its period increases, as soon as $\alpha$ increases
past the threshold value of $\pi/2$.
\begin{figure}
\resizebox{0.26\textwidth}{!}{\includegraphics{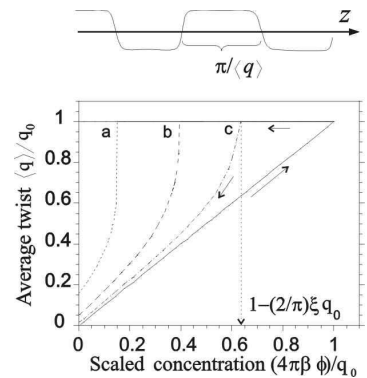}}
\caption{Relative helical wave number as function of chiral dopant
fraction $\phi$ for the MW model. On increasing $\phi$ the wave
number increases; the crosslinking occurs at $q=q_0$. On
subsequent decreasing $\phi$ the strength of imprinting depends on
parameter $\xi q_0$:  curves ({\sf a}) $\xi q_0=0.75$, ({\sf b})
$\xi q_0=1.05$, ({\sf c}) $\xi q_0=1.75$.  } \label{models}
\end{figure}

Fig.~\ref{models} illustrates the model results by plotting the
ratio $\langle q \rangle / q_{0}$, which is the relative number of
remaining cholesteric phase inversions (helix periods). After
crosslinking at $\phi = \phi_{0}, q=q_0$, on removing the chiral
dopant the network initially is not affected, until a critical
value $4\pi\beta \, \phi^* = q_{0}-2/(\pi\xi)$ is reached. This
critical point may or may not be accessed on de-swelling,
depending on the values of $\xi$ and $q_0$. The value of $\langle
q \rangle$ still remaining at $\phi=0$ is the amount of
topological imprinting by the network. Very low crosslink density
leads to low $D_1$, high $\xi$ and the nearly complete unwinding
of helices (loss of imprinting). A highly crosslinked network,
leads to low $\xi$ and, if $\phi^* \le 0$, the complete retention
of the original helix.

It appears that the average pitch wave number $\langle q \rangle$
is the pertinent variable. The theoretical model have its strong
and weak points, but the underlying physical effect is clear and
universal. Our first task is to experimentally study the process
of topological imprinting of chirality. Its measurement is
possible due to the remarkable optical property of cholesterics:
their giant optical rotation. The eigensolutions for
electromagnetic waves propagating along $z$ are superpositions of
two circularly polarized plane waves of opposite signs
\cite{deVries:51}. Their phase difference gives the optical
rotation $\Psi$ in the medium, the rate of which can be related to
the average pitch wave number $\langle q \rangle$ by the de~Vries
equation:
\begin{equation}
\frac{d\Psi}{dz} =- \langle q \rangle a^{2} \left[ 4 \left( \frac{
\langle q \rangle \Lambda}{\pi\sqrt{\bar{\varepsilon}}}
\right)^{2} \left(1 - \left( \frac{ \langle q \rangle
\Lambda}{\pi\sqrt{\bar{\varepsilon}}} \right)^{2} \right)
\right]^{-1} , \label{optical}
\end{equation}
where $\Lambda$ is the wave length of the incident light (He-Ne
laser, $633 \, \hbox{nm}$), $a= (n_{e}^{2}-n_{o}^{2})/
(n_{e}^{2}+n_{o}^{2})$ is the dielectric anisotropy parameter
(where in our material the extraordinary and ordinary refractive
indices are, respectively, $n_{e}=1.75$ and $n_{o}=1.6$) and in
zero-order approximation $\bar{\varepsilon} =\frac{1}{2}
(n_{e}^{2}+n_{o}^{2})$ is the mean dielectric constant. For
$\langle q \rangle =\pi\sqrt{\overline{\varepsilon}}/\Lambda$ a
dispersion anomaly appears in the form of Bragg-like reflection
\cite{deVries:51}. In our case this anomaly would occur at
$\langle q \rangle/\pi \approx (381\, \hbox{nm})^{-1}$. In
practice, we obtain the initial wave number of cholesteric state
$q_{0}/\pi \approx (496 \, \hbox{nm})^{-1}$, which means that our
range of measurements is always on the longer wavelength side of
the point of divergence in Eq.~(\ref{optical}).

Experimentally, the optical rotation $\Psi$ can be determined
using a dynamical method based on measuring the phase difference
between the split parts of linearly polarized laser beam, one
passing through the sample and the rotating analyzer, the other
through the chopper (providing the reference signal to lock on).
The phase between the two beams is measured by an integer number
of periods and corresponds directly to the optical rotation
$\Psi$, from which the parameter $\langle q \rangle$ is then
calculated.

\begin{figure}
\resizebox{0.37\textwidth}{!}{\includegraphics{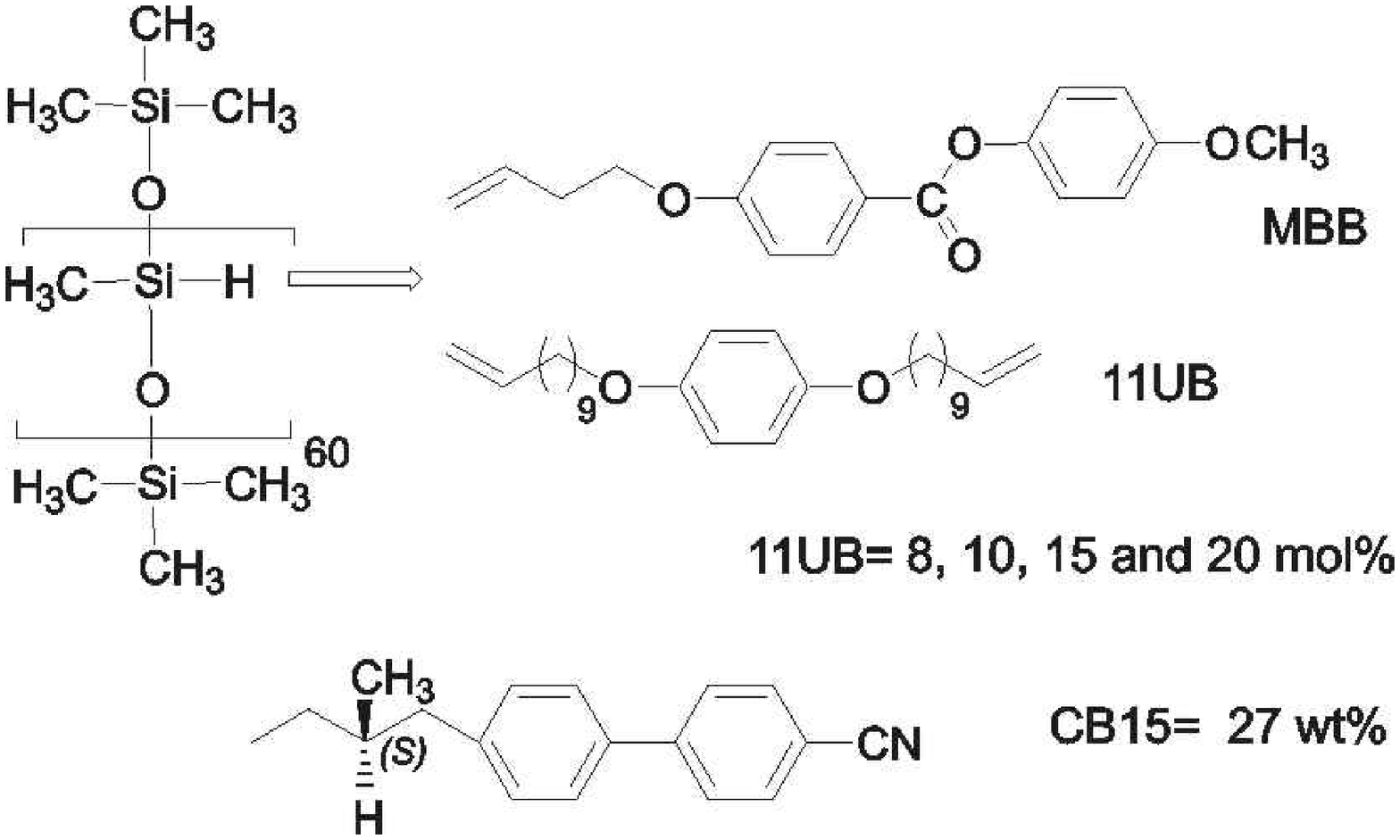}}
\caption{Chemical composition of the elastomer network. }
\label{ali}
\end{figure}
\begin{figure}
\resizebox{0.47\textwidth}{!}{\includegraphics{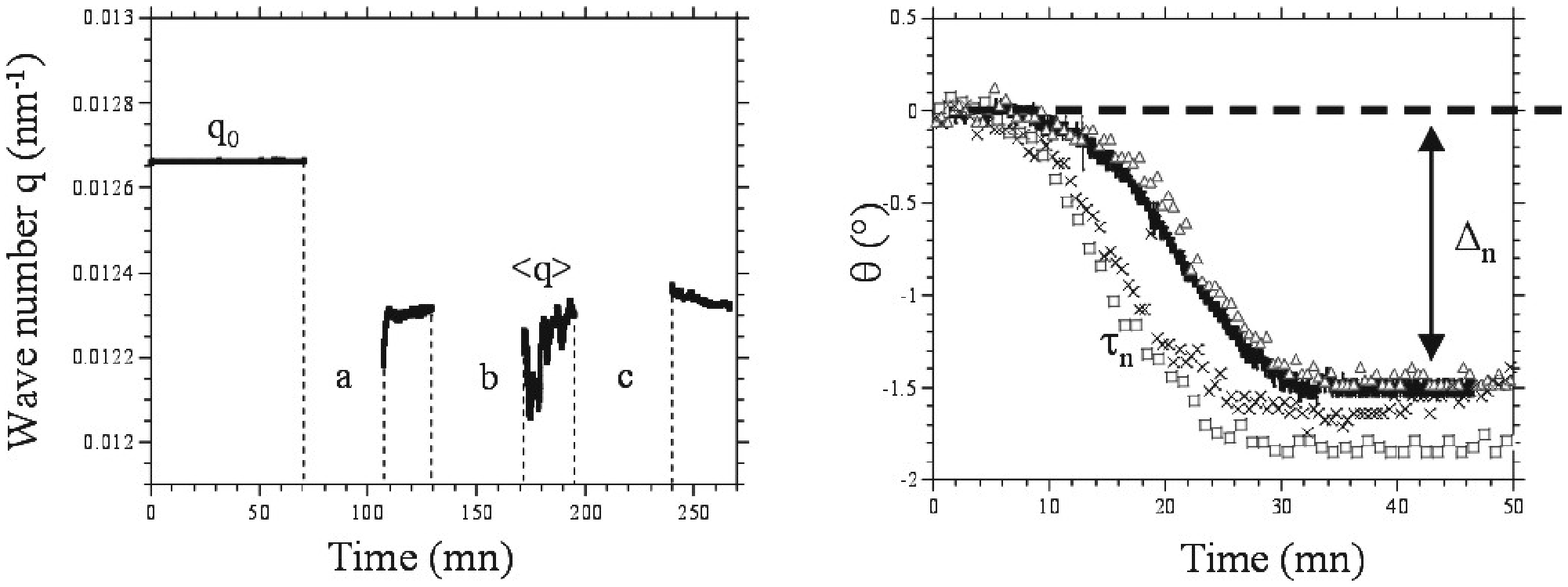}}
 \caption{(a) -- Chiral dopant release from the imprinted
network into the ``wash''.  \ (b) -- The effect of CB15 removal on
the inverse helical pitch for the 10\%-crosslinked network.  }
\label{imprinting}
\end{figure}

From the theory, we have seen that the robustness of chiral
imprinting depends on the nematic penetration depth $\xi=
\sqrt{K_{2}/D_{1}}$, which can be controlled by varying the
density of crosslinks in the network (affecting $D_{1}$). The
preparation of imprinted polysiloxane cholesteric elastomers
follows the pioneering work of Kim and Finkelmann \cite{Kim:01},
which obtains monodomain cholesteric elastomers by uniaxial
de-swelling during crosslinking. (Monodomain textures, with the
cholesteric pitch uniformly aligned along the optical path, are
essential for the study of giant optical rotation). The mesogenic
group (4'-methoxyphenyl 4-(buteneoxy)benzoate, MBB) and the
crosslinker (1,4-di(11-undeceneoxy)benzene, di-11UB), both
synthesized in-house, with the molar ratios of 9.2:0.8, 9:1,
8.5:1.5 and 8:2, doped with a fixed concentration (27\% of total
weight) of chiral compound (4-(2-methylbutyl)-4'-cyanobiphenyl,
CB15), from Merck, were reacted with polymethylsiloxane chains in
toluene, Fig.~\ref{ali}. After evaporation of the solvent and
completion of crosslinking, cholesteric elastomers were obtained
-- with the same helical pitch $\pi/q_0$, irrespective of the
crosslinking density. In order to remove the chiral dopant CB15,
the material is placed in a large volume of non-chiral solvent
(acetone) leading to a diffusion of CB15 from the network in
response to a concentration gradient. The release of the chiral
dopant is then determined by measuring the rotation of the
polarization plane $\widetilde{\Psi}$ of the bulk solvent with a
differential detection: $\widetilde{\Psi} = \arcsin
(\frac{E_{x}-E_{y}}{E\sqrt{2} })$ where $E_{x,y}$ is the signal
amplitude for the two perpendicular polarizations.
Figure~\ref{imprinting}(a) shows the evolution of
$\widetilde{\Psi}$ for elastomers with different crosslink density
``washed'' in a non-chiral solvent. For all materials, the value
of $\widetilde{\Psi}$ increases and saturates for
$\widetilde{\Psi}=-1.5$~deg (for this ordinary isotropic Faraday
effect $\widetilde{\Psi}=\varphi d \beta$, where $\varphi$ is the
solute concentration, $d$ the optical path and $\beta$ the
twisting power of CB15). The characteristic time of dopant release
is approximately the same for all elastomers and is $\tau \sim
20$min, at room temperature, showing perhaps a slight increase
with crosslink density.

\begin{figure}
\resizebox{0.47\textwidth}{!}{\includegraphics{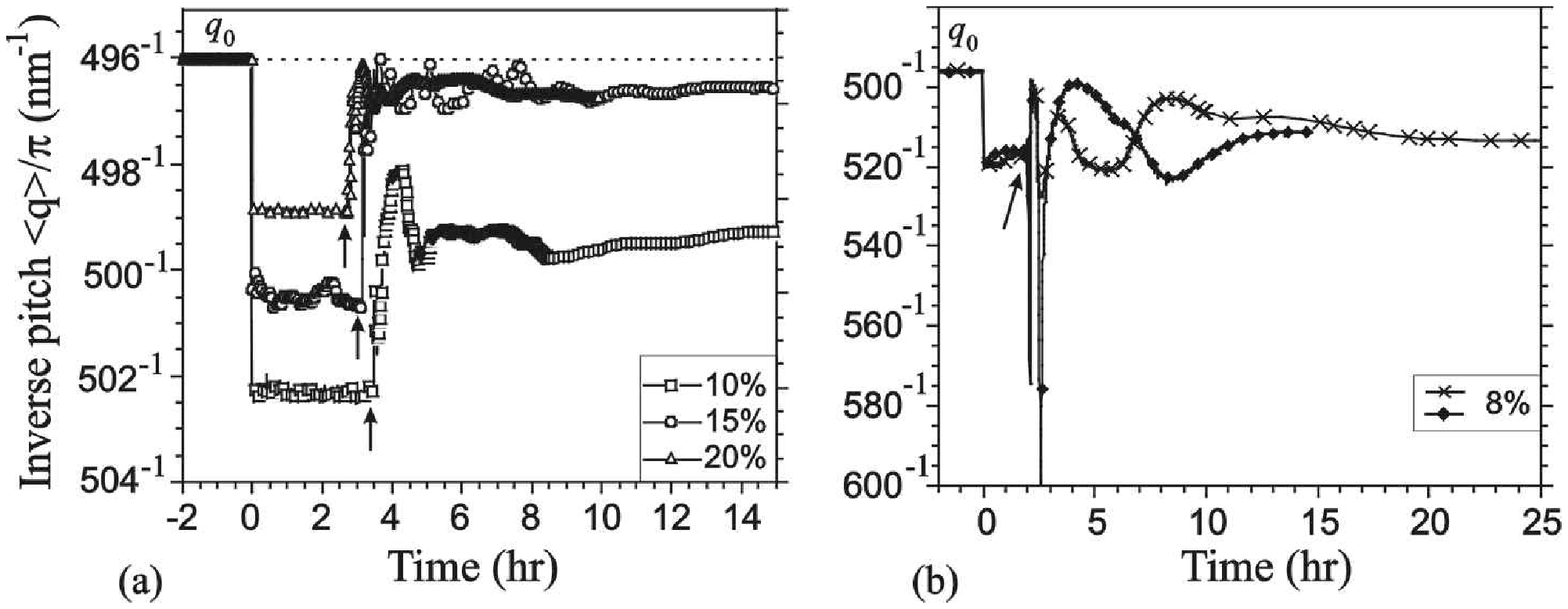}}
 \caption{The inverse cholesteric pitch of imprinted elastomers,
$p^{-1}=\langle q \rangle/\pi$. The plots show the initial value
at crosslinking, the imprinted plateau level of the dry network,
and the change after exposure to racemic mixture. (a) -- Results
for 10\%, 15\% and 20\% crosslink density, (b) -- for 8\%
crosslinked network. Arrows represent the moment of deposition of
the racemic droplet onto the network. } \label{imprinted swelling}
\end{figure}

After drying, does the resulting elastomer still retain a residual
macroscopic (phase) chirality with only centrosymmetric molecules
left in the network? Figure~\ref{imprinting}(b) is the case for
the network with 10\% crosslinking, which shows that the average
wave number remains non-zero after the ``wash'',  $\langle q
\rangle/\pi \approx (505\, \hbox{nm})^{-1}$, although lower than
the initial value in the presence of chiral dopant. Repeated
flushes with acetone evidently do not have any further effect. We
find the same initial wave number ${q_{0}}/\pi \approx (496 \,
\hbox{nm})^{-1}$ for all elastomers, but the amount of retained
phase chirality is a function of the crosslink density, see
Fig.~\ref{imprinted swelling}. As theory predicts, the remaining
fractional wave number $\langle q \rangle/q_0$ increases as the
parameter $\xi q_0$ decreases and the director $\bm{n}$ is more
strongly anchored to the network.

Now if the imprinted elastomer is brought in contact with a
racemic mixture (a 50/50 proportion of left- and right-handed
small molecules, 2-Bromopentane, from ACROS), its natural wave
number $q_{0}$ can be restored by selective absorption of chiral
molecules which agree with the imprinted handedness of the
network, and rejecting the molecules with wrong handedness. The
experiment is straightforward: the imprinted network is swollen in
the racemic solvent, and then allowed to dry again. With a
stereo-neutral solvent, nothing happens with the network, as the
repeated cycles in Fig.~\ref{imprinting}(b) indicate.
Figure~\ref{imprinted swelling} shows that densely crosslinked
networks (15 and 20\%) are able to retain a sufficient amount of
left-handed (S) component of the mixture to return their average
cholesteric pitch to nearly the same level as they had before the
CB15 removal: $496.9$ and $496.6\, \hbox{nm}$, respectively.
Networks with lower crosslinking density show less
stereo-selectivity. Elastomers with weaker crosslinking retain
much less of the racemic solvent, returning their pitch to $499$
for 10\% and to $512\, \hbox{nm}$ for 8\% network. Note that in
both 10 and 8\%-crosslinked elastomers one also finds great
instability and high amplitude of alternating pitch as the network
becomes increasingly weaker. This effect is not experimental
noise, but rather reproducible oscillations with amplitude
decreasing and period increasing with time, cf. two different
curves in Fig.~\ref{imprinted swelling}(b). In fact, we observe
similar oscillations in many situations of de-swelling liquid
crystalline elastomers. Although no full theoretical explanation
exists, we attribute this effect to non-uniform distribution (in
space and time) of coupled solvent density and local nematic
order.
\begin{figure}
\resizebox{0.3\textwidth}{!}{\includegraphics{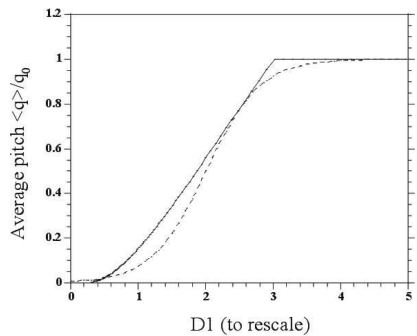}}
 \caption{The comparison between the values of inverse pitch in the
imprinted networks of different crosslink density ($\times$) and
the weight percent of the amount of (S)-component extracted by
these networks from a racemic mixture ($\circ$). }
\label{percents}
\end{figure}

The physical trend is clear: at low crosslinking density, the
selectivity is weak; highly crosslinked networks simply cannot
swell very much. How can we quantify these measurements, which
unambiguously demonstrate the stereo-selectivity of imprinted
cholesteric networks? Directly applying the model \cite{Mao:01b}
is impossible because even a small proportion of non-mesogenic
dopant reduces the local order parameter and thus affects
practically all model parameters. An extension of theory is
necessary to incorporate these effects, which is a part of a
forthcoming publication, also presenting a number of experimental
details (such as the variation of order parameter $Q(T,\phi)$,
swelling kinetics and inhomogeneous network deformations on
swelling/drying). Here we simply quote the measurements of the
total weight intake of the solvent. Assuming that only the
left-handed component of the solvent is retained (nothing to stop
the molecules with ``wrong'' chirality to evaporate), the figures
give an idea of the stereo-separation efficiency. The
20\%-crosslinked imprinted network retains 5.7 wt\% of
2(S)-Bromopentane; the 15\% network -- 5.3 wt\%, the 10\% -- 4.7
wt\% and the 8\% network -- 4 wt\%, Fig.~\ref{percents}.

In Summary, we have experimentally demonstrated the topological
imprinting of phase chirality in liquid crystalline polymer
networks. The imprinting efficiency depends on the chiral order
parameter $\alpha = \xi q_0$, a function of elastic constants
$K_{2}$ and $D_{1}$. A spectacular property of imprinted networks
is their capacity to preferentially absorb and retain right- or
left- molecules from a racemic solvent. The subsequent release of
the chiral solvent component is easily achieved by bringing the
cholesteric elastomer into the isotropic phase, when no phase
chirality remains to provide the bias, or alternatively -- by
stretching the rubber, which results in the helix unwinding. We
did not demonstrate this explicitly in this short Letter (although
Fig.~\ref{imprinting}(b) gives an indication of reversibility). It
should suffice to reassure the reader that all stereo-selectivity
results were completely reproducible and reversible, after a
sample was annealed into the isotropic phase. This makes the
described phenomenon a practical possibility for many
applications.

We thank Y. Mao and M. Warner for helpful discussions and
acknowledge financial support from EPSRC.


\end{document}